\documentclass[useAMS,onecolumn]{mn2e}

\def\be{\begin{equation}}
\def\ee{\end{equation}}
\def\bea{\begin{eqnarray}}
\def\eea{\end{eqnarray}}
\def\etal{{\it et al.~}}

\title[]{Polarization evolution accompanying the very early sharp decline of GRB X-ray afterglows}
\author[]{Yi-Zhong Fan$^{1,2,3}$\thanks{E-mail: yzfan@pmo.ac.cn},
Dong Xu$^4$ and Da-Ming Wei$^{1,2,5}$\\
$^1${\sl Purple Mountain Observatory, Chinese Academy of Sciences, Nanjing 210008, China}\\
$^2${\sl National Astronomical Observatories, Chinese Academy of Sciences, Beijing 100012, China}\\
$^3${\sl Niels Bohr Institute, Neils Bohr International Academy, Blegdamsvej 17, DK-2100 Copenhagen, Denmark}\\
$^4${\sl Dark Cosmology Centre, Niels Bohr Institute, University of Copenhagen, Juliane Maries Vej 30, 2100 Copenhagen, Denmark}\\
$^5${\sl Joint Center for Particle Nuclear Physics and Cosmology of
Purple Mountain Observatory - Nanjing University, Nanjing 210008,
China.}\\}
\date{Accepted 2007 December 19. Received 2007 December 04; in original form 2007 October 02}

\begin{document}

\maketitle
\begin{abstract}
In the synchrotron radiation model, the polarization property
depends on both the configuration of the magnetic field and the
geometry of the visible emitting region. Some peculiar behaviors in
the X-ray afterglows of {\it Swift} gamma-ray bursts (GRBs), such as
energetic flares and the plateau followed by a sharp drop, might by
highly linearly-polarized because the outflows powering these
behaviors may be Poynting-flux dominated. Furthermore, the
broken-down of the symmetry of the visible emitting region may be
hiding in current X-ray data and will give rise to interesting
polarization signatures. In this work we focus on the polarization
accompanying the very early sharp decline of GRB X-ray afterglows.
We show that strong polarization evolution is possible in both the
high latitude emission model and the dying central engine model
which are used to interpret this sharp X-ray decline. It is thus not
easy to efficiently probe the physical origin of the very early
X-ray sharp decline with future polarimetry. Strong polarization
evolution is also possible in the decline phase of X-ray flares and
in the shallow decline phase of X-ray light curves characterized by
chromatic X-ray VS. Optical breaks. An {\it XRT}-like detector but
with polarization capability on board a {\em Swift}-like satellite
would be suitable to test our predictions.

\end{abstract}

\begin{keywords}
Gamma Rays: bursts $-$ polarization $-$ GRBs: jets and outflows $-$
radiation mechanisms: nonthermal
\end{keywords}

\section{Introduction}
\label{sec:Observation} The successful launch of the {\it Swift}
satellite in Nov 2004 opened a new window to reveal what has been
happening in the early afterglow phase of gamma-ray bursts (GRBs).
As summarized in Zhang et al. (2006) and Nousek et al. (2006), in a
canonical {\it Swift} GRB X-ray afterglow lightcurve some
interesting features are emerging (detected in a good fraction of
but not all bursts), including a very early sharp decline (phase-I),
a shallow decline or even plateau, (phase-II), preceding the
conventional or so-called ``normal" decay phase (phase-III \& -IV),
and the energetic X-ray flares (phase-V). Various modifications of
the standard GRB afterglow model have been put forward to explain
the observations (see M\'esz\'aros 2006; Piran \& Fan 2007; Zhang
2007 for recent reviews). Although the underlying physical processes
are not very clear, it is widely suggested that the X-ray Phases I,
II and V may be relevant to prolonged activities of the central
engine while the simultaneous optical afterglow is dominated by GRB
forward shock emission (Fan \& Wei 2005; Zhang et al. 2006;
Ghisellini et al. 2007; Panaitescu 2007). If so, the early time
UV/optical and X-ray polarization behaviors should be independent
and may be very different from each other. In some models, the
outflows powering energetic flares and X-ray plateaus followed by a
sharp drop are Poynting-flux dominated (e.g., Gao \& Fan 2006;
Giannios 2006; Troja et al. 2007), thus high-linear polarization of
these synchrotron radiation photons is expected (Fan, Zhang \& Proga
2005).

Except the configuration of the magnetic field, the geometry of the
emitting area can influence the polarization property of the
synchrotron radiation as well. The broken-down of the symmetry of
the visible emitting region may be hiding in the peculiar X-ray data
and can give rise to interesting polarization signatures. Motivated
by this idea, in section 2 we calculate the polarization property of
Phase-I and show that strong evolution is likely. In section 3, we
argue that similar phenomena are also expected in the decline phase
of the X-ray flares and possibly in the late time part of Phase-II,
especially those associated with a chromatic break in optical band.
We also discuss the detection prospect there. In section 4, we
summarize our work with some discussions.

\section{The polarization evolution accompanying Phase-I}
For phase-I, two leading interpretations are: (i) these X-ray
photons are actually the high latitude emission of a main internal
shock pulse (Fenimore et al. 1996; Kumar \& Panaitescu 2000; Zhang
et al. 2006) while the central engine has died; and (ii) these X-ray
photons are powered by the weaker and weaker activity of the central
engine, i.e., the sharp X-ray decline traces the activity of the
dying central engine (Fan \& Wei 2005; Piran \& Fan
2007)\footnote{These two interpretations can be also applied to the
sharp decline of the X-ray flares.}. The interpretation (i) has been
widely accepted because some X-ray declines can be adjusted as
$\propto (t-t_0)^{-(2+\beta_{\rm X})}$ predicted by the high
latitude emission model (Liang et al. 2006; Yamazaki et al. 2006;
Zhang, Liang \& Zhang 2007a), where $\beta_{\rm X}$ is the X-ray
spectral index, and $t_0$ is a free parameter parameterizing the
beginning of a certain internal shock pulse \footnote{In this
interpretation, the duration of the main prompt emission pulse
should be in order of the timescale of whole burst otherwise the
total high latitude emission of those early pulses decay too fast to
account for the observation (see Fig.2 of Fan \& Wei 2005 for
illustration). This is somewhat unusual in the standard internal
shock model, in which the typical variability timescale is just in
order of milliseconds. A possible solution of this puzzle is to
assume that the prompt $\gamma-$ray emission are powered by some
magnetic energy dissipation processes at a distance $R_{\rm
prompt}\sim 10^{15}-10^{16}$ cm to the central engine, so the
emission duration is governed by the angular timescale $\sim R_{\rm
prompt}/(2\Gamma^2 c) \sim 6 (R_{\rm prompt}/10^{15}~{\rm
cm})(\Gamma/50)^{-2}$ sec, where $\Gamma$ is the bulk Lorentz factor
of the outflow.}. Such a test, however, is not conclusive. As shown
in Sakamoto et al. (2007), for the very sharp X-ray decline, the
data itself can not well constrain the function form $(t-t_0)^{-k}$.
In many events, $k$ can be taken as a constant (for example, $k \sim
2$) and thus is not relevant to $\beta_{\rm X}$. Additionally, it is
not easy to see why the GRB central engine turns off abruptly while
the numerical simulation actually shows the contrast (MacFadyen et
al. 2001; Zhang, Woosley \& Heger 2007b). Interpretation (i) also
fails to interpret some slowly decaying phase-I as in GRB 060614
\cite{BK07,ZLZ07,Xu07}. As for interpretation (ii), there has been
no strict observational test so far though it seems possible for any
energetic burst phenomenon.

We take into account both models in our calculation and examine the possibility to distinguish
between these two models with future X-ray polarimetry.

\subsection{High latitude emission model}
If interpretation (i) is correct, i.e., the sharp X-ray decline is due to the high latitude
emission of a main internal shock pulse, there would be very interesting polarization signals.
As is known, GRB outflow is likely to be jetted. In this work we only consider the uniform jet
model. Following Ghisellini \& Lazzati (1999), we assume that the half-opening angle of the
ejecta is $\theta_{\rm j}$, and that the line of sight (L.o.S.) makes an angle $\theta_{\rm
v}$ with respect to the jet's central axis (C.A., see Fig.1a for illustration). The
probability of observing the ejecta along its C.A. is vanishingly small since it corresponds a
very small solid angle. For typical bright GRBs, the L.o.S. is likely to be within the cone,
i.e., $\theta_{\rm v}<\theta_{\rm j}$. At early time, the high latitude emission is from zones
around the L.o.S. satisfying $\theta \leq \theta_{\rm j}-\theta_{\rm v}$, so the net
polarization of the detected photons vanishes because of the symmetry; but at later time, the
high latitude emission is from $\theta>\theta_{\rm j}-\theta_{\rm v}$, the symmetry is broken
and the observed net polarization is not zero any longer.

\begin{figure}
\begin{picture}(0,200)
\put(0,0){\includegraphics{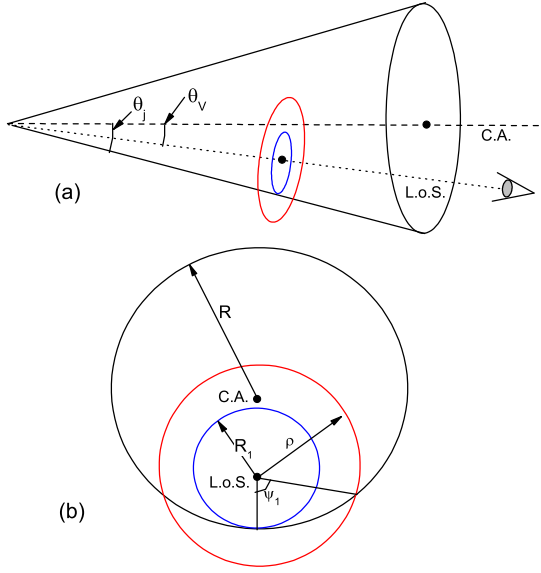}}
\end{picture}
\caption{(a) The high latitude emission model for the X-ray decline. The larger the angle with
respect to the line of sight (L.o.S), the latter the emission arrives at us because of the
geometry effect. (b) Sketch of the geometrical set-up used to compute the polarization signal
(see also Ghisellini \& Lazzati 1999).} \label{fig:Cartoon}
\end{figure}

In this model, the photons arrived at $t$ were emitted from
\begin{equation}
\theta \approx (2 c t/R_0)^{1/2},
\end{equation}
where $R_0 \sim R_{\rm prompt}$ is the radius of the main internal
shock pulse and $c$ is the speed of light. Below we adopt the
simplified calculation of Ghisellini \& Lazzati (1999). The
notations used in Fig.\ref{fig:Cartoon}b are related to $R_0$,
$\theta_{\rm j}$ and $\theta_{\rm v}$ as $R\approx R_0 \sin
\theta_{\rm j}$ and $R_1\approx R_0 \sin(\theta_{\rm j}-\theta_{\rm
v})$. When $\rho>R_1$, the ring is not a complete circle. The
missing parts are in $(0,~\psi_1)$ and $(2\pi-\psi_1,~2\pi)$ as
shown in Fig.\ref{fig:Cartoon}. Approximately, we have (Ghisellini
\& Lazzati 1999)
\begin{equation}
\psi_1 \approx
\left\{%
\begin{array}{ll}
    {\pi \over 2}-\arcsin[{2R_1R-R_1^2-\rho^2 \over 2\rho
(R-R_1)}], & \hbox{for $\rho>R_1$;} \\
    0, & \hbox{for $\rho<R_1$,} \\
\end{array}%
\right.
\end{equation}
where $\rho \approx R_0 \sin \theta$.

In principle, the GRB outflow could be either baryon-rich or
Poynting-flux dominated. For the former the magnetic field is
generated in the shock front and is likely to be random. For the
latter the magnetic field is coherent in a large scale. The
resulting polarization light curves are expected to be different in
these two cases, as shown below.

{\bf Random magnetic field.} The net polarization of the emission
from $\rho - \rho+d\rho$  can be approximated by (See the Appendix
for the derivation)
\begin{eqnarray}
\Pi(\rho) &\approx & \Pi_0{(2\zeta^2-1)\sin^2 \alpha \over 2 \zeta^2
\sin^2 \alpha +(1+\cos^2 \alpha)}{\mid \sin(2\psi_1) \mid \over
2\pi-2 \psi_1},
\label{eq:Main}
\end{eqnarray}
where $\Pi_0$  is the maximum polarization at one point, $\alpha$ is
the viewing angle with respect to the fluid velocity in the outflow
comoving frame (see Figure \ref{point} for illustration), which is
governed by $\cos \alpha =(\cos \theta-\beta)/(1-\beta \cos
\theta)$, and the parameter $\zeta^2=<B_\parallel^2>/<B_\bot^2>$
denotes the level of anisotropy of the magnetic field distribution,
where $B_\perp$ and $B_\parallel$ are magnetic field perpendicular
and parallel to the normal of the shock plane.
$\beta=(1-1/\Gamma^2)^{1/2}$ is the velocity of the GRB ejecta in
units of $c$.

We note that  in eq.(\ref{eq:Main}) $\Pi(\rho) \propto
\sin^2\alpha/(1+\cos^2\alpha)$ if $\zeta^2 \sim 0$. For $\Gamma
(\theta_j-\theta_v)\gg 1$, we have $\Pi(\rho) \propto 2\Gamma^2
(\theta_{\rm j}-\theta_{\rm v})^2/[1+\Gamma^4 (\theta_{\rm
j}-\theta_{\rm v})^4]\ll 1$. The resulting net polarization can thus
be neglected. But for $2\zeta^2 \sin^2 \alpha \gg 1+\cos^2\alpha$,
the second term in the right side of eq.(\ref{eq:Main}) is
$\rightarrow 1$. Significant net polarization is possible.

As $\theta>(\theta_{\rm j}-\theta_{\rm v})$, the observers just
detected $\sim (\pi-\psi_1)/\pi$ part of the circle, the X-ray flux
will decline more steeply than $t^{-(2+\beta_{\rm X})}$  and is
governed by
\begin{equation}
F_X(t) \propto {\pi-\psi_1(t) \over \pi} t^{-(2+\beta_X)}.
\end{equation}

\begin{figure}
\begin{picture}(0,240)
\put(0,0){\includegraphics{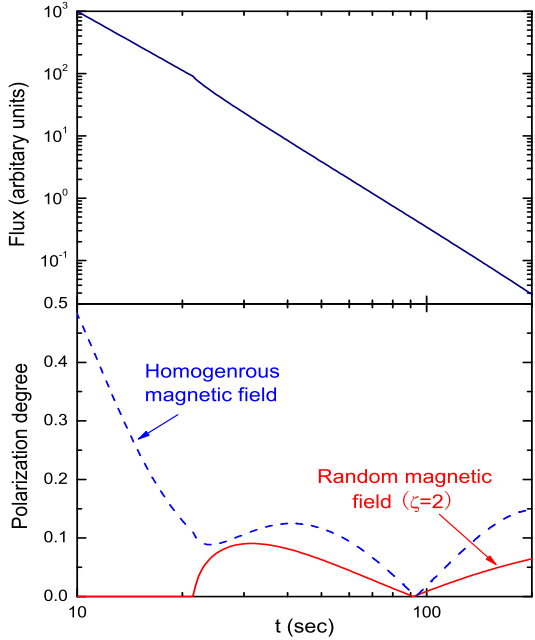}}
\end{picture}
\caption{The sharp X-ray decline (upper panel) and the corresponding polarization light curves
(lower panel) after a time-shift correction, in the case of the high latitude emission model.
For this figure we assumed $\Pi_0=60\%$, $\Gamma=50$, $\theta_{\rm j}=0.1$, $\theta_{\rm
v}=0.07$,
 $R_0=10^{15}$ cm and $\beta_X=1.15$.} \label{fig:LC}
\end{figure}

We plot in Fig.\ref{fig:LC} the sharp X-ray decline and the corresponding polarization light
curves. When the emission is from a circle ring, the net polarization is zero because of the
symmetry. As $\theta$ is getting larger and just part of the ring is visible, the observed
polarization degree is not zero any longer until $\rho^2 \rightarrow 2R_1R-R_1^2$ (i.e.,
$\psi_1 \rightarrow \pi/2$). At the point of $\psi_1=\pi/2$, the net polarization becomes zero
again according to equation (3), because the polarization direction abruptly changes by
$\pi/2$. After this point, the polarization increases again. But the flux may be too weak to
have a significant polarization measurement.

{\bf Homogenous magnetic field.}  In this case, the instantaneous
stokes parameters (note that \( V=0 \) as the polarization is
linear) and the polarization degree are given by (Granot \& K\"onigl
2003):
\begin{equation} \label{Eq QU ordered} \frac{\left\{ \begin{array}{c}Q \\U \\
\end{array}\right\}}{I}=\Pi _0 \frac{\int _{\psi_1}^{2\pi-\psi_1 }
 [({1-y \over 1+y})^2 \cos^2\phi+\sin^2 \phi]^{\epsilon/2}
\left\{ \begin{array}{c} \cos(2\theta _{p}) \\ \sin(2\theta _{p}) \\
\end{array}\right\} d\phi }{\int _{\psi_1}^{2\pi-\psi_1}[({1-y \over 1+y})^2
\cos^2\phi+\sin^2 \phi]^{\epsilon/2}d\phi},
\end{equation}
and
\begin{equation} \label{Eq Pi} \Pi
=\frac{\sqrt{U^{2}+Q^{2}}}{I},
\end{equation}
where \( \theta _{p}=\phi +\arctan(\frac{1-y}{1+y}\cot\phi ) \),
$y=\Gamma^2\theta^2$ and $\epsilon=(1+\beta_X)/2$ (Granot \&
K\"onigl 2003). Again the polarization light curve has been plotted
in Fig.\ref{fig:LC}, which is similar to that of the random magnetic
field case except an initial high polarization degree. This strongly
suggests that the late time non-zero polarization is mainly
contributed by the geometry effect, as in the case of a random
magnetic field. If the symmetry has not been broken down, the late
time net polarization for an ordered magnetized outflow disappears,
as already found in Nakar, Piran \& Waxman (2003).

\subsection{Dying central engine model}
\begin{figure}
\begin{picture}(0,200)
\put(0,0){\includegraphics{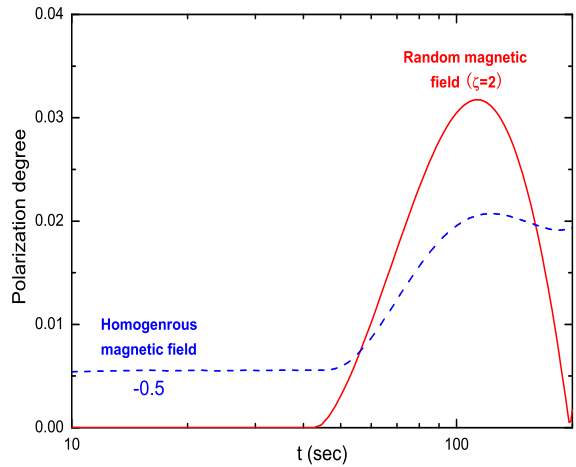}}
\end{picture}
\caption{The polarization light curves in the case of dying central
engine model. The physical parameters are the same as that of
Fig.\ref{fig:LC} except we assume $\Gamma=150(t/10)^{-1}$ for
$10<t<200$ sec. Please note that the dashed line is not the real
polarization degree unless a factor 0.5 has been added.}
\label{fig:LC2}
\end{figure}

Generally, in the dying central engine model there are two
possibilities. (1) If the outflow powering the fast decaying X-ray
tail emission always has a Lorentz factor $\Gamma>1/(\theta_{\rm
j}-\theta_{\rm v})$, no strong polarization evolution is expected
unless the physical composition of the outflow has changed a lot
(e.g., from byron dominated to Poynting-flux dominated). We thus may
be able to distinguish between the high latitude emission model and
the dying central engine model with the future X-ray polarimetry.
(2) The bulk Lorentz factor of the outflow drops so quickly as to
satisfy $\Gamma< 1/(\theta_{\rm j}-\theta_{\rm v})$ in a short
period. When $\Gamma<1/(\theta_{\rm j}-\theta_{\rm v})$, the
symmetry of the visible emitting area is broken and interesting
polarization signature begins to exist. Since there is little
information about the evolution of $\Gamma$ with $t$, we here simply
assume $\Gamma \approx 150(t/10)^{-1}$ for $10<t<200$ sec and
$\theta_{\rm j}$ and $\theta_{\rm v}$ are constant. Again, we
calculate the polarization evolution in the cases of {\it random
magnetic field} and {\it Homogenous magnetic field}. But now the
calculation should be due to the emission area within $\theta \leq
1/\Gamma$, rather than from a ring at $\theta$ in the high latitude
emission model. Therefore, we calculate the quantities of
$(Q,~U,~I)$ along the radius $\rho$ and integrate them over $\rho$.
Such calculations have been carried out by Ghisellini \& Lazzati
(1999) and Granot \& K\"onigl (2003), respectively. Our numerical
results are presented in Figure \ref{fig:LC2}. One can see that
polarization evolution is evident and shares some similarity with
that in Fig.\ref{fig:LC}, particularly in the case of {\it random
magnetic filed}, though the polarization degree of high latitude
emission model is larger than that of the dying central engine
model. This is reasonable since in the former case the emission is
from the ring while in the latter case the emission is from
$\theta\leq 1/\Gamma$. In the case of {\it homogenous magnetic
filed}, the evolution is small because of the significant
polarization background (the degree is $\sim 0.5$) caused by the
ordered magnetic field from the central source.

\section{Prospect of detecting the X-ray polarization}
Our calculations in Section 2 are actually applicable to more phases in the GRB X-ray
afterglow light curve. One is the decline phase of X-ray flares (in Phase-V), which is also
attributed to the high latitude emission model or the dying (re-)activity of the central
engine. The other is the late phase of the X-ray flattening (in Phase-II). Recently Ghisellini
et al. (2007) interpreted Phase-II, particularly those show chromatic breaks in X-ray and
optical bands, as the prolonged emission of the central engine. The X-ray flattening ceases to
exist when the bulk Lorentz factor of the outflow is so small that one sees the edge of the
outflow. If these interpretations are correct, strong polarization evolution should be present
at the end of Phase-II and in the decline of Phase-V.

Measuring polarization is of growing interest in high energy astronomy. New technologies are
being invented, and several polarimeter projects are under construction. In the X-ray band,
the ongoing projects include XPE (Elsner et al. 1997), PLEXAS (Marshell et al. 1998), POLAR
(Produit et al. 2005), XPOL (Costa et al. 2006), POET (Polarimeters for Energetic Transients,
see McConnel et al. 2007), and so on. For example, the planned POET includes two instruments:
GRAPE (Gamma-Ray Polarimeter Experiment) and LEP (Low Energy Polarimeter, working in soft
X-ray band). So it may be suitable to detect the polarization signatures predicted in the
early X-ray afterglow of GRBs. An important issue is whether any of these detectors could
perform a prompt slew to the GRBs localized by the $\gamma-$ray monitor. An ideal instrument
would be an XRT-like detector with polarization capability on board a {\em Swift}-like GRB
mission.

\section{Summary \& Discussion}
Many surprises in GRB afterglows, mainly in the X-ray band, have been brought since the
successful launch of {\it Swift} in 2004. Though not clearly understood yet, the (re-)activity
of GRB central engine is likely to play an important role on producing afterglows, as
speculated by Katz, Piran \& Sari (1998). The different temporal behaviors in the UV/optical
and X-ray afterglow favor that the peculiar X-ray emission traces the prolonged activity of
the central engine while the optical afterglow is dominated by the forward shock emission
\cite{FW05,GG07}. If true, the polarization behaviors of the early X-ray and optical afterglow
may be independent and different. At such an early time, the linear polarization of the
optical forward shock emission is expected to be very small \cite{GL99} unless the
interstellar medium is magnetized \cite{GK03}, because the Lorentz factor of the outflow is so
large that the visible emitting region is well within the cone of the ejecta and is thus
symmetric. But in the X-ray band, interesting polarization signatures may be present. For
example, Fan et al. (2005) argued that some X-ray flares might have a high linear polarization
because these new outflows are likely to be Poynting-flux dominated. Such an argument is also
applicable to the X-ray plateau followed by a sharp drop as detected in GRB 070110.

In this work, we focus on the polarization properties of the very
early sharp decline of X-ray afterglows (Phase-I). In the high
latitude emission model, the later the emission, the larger the
angular $\theta$ (see Figure \ref{fig:Cartoon} for details). So at
late time, the symmetry of the emitting ring is broken when part of
the emitting ring is out of the cone of the ejecta. Strong
polarization evolution thus emerges (see Section 2.1). For
comparison, results from the dying central engine model are a bit
more complicated. If the outflow powering the fast decaying X-ray
tail emission always has a Lorentz factor $\Gamma>1/(\theta_{\rm
j}-\theta_{\rm v})$, no strong polarization evolution is expected
unless the physical composition of the outflow changed
significantly. However, if the Lorentz factor drops rapidly to
satisfy $\Gamma<1/(\theta_{\rm j}-\theta_{\rm v})$, then
polarization evolution is evident. In this scenario, if the magnetic
field involved is shock-generated and is thus random, then strong
polarization evolution is expected; while if the magnetic field
involved is brought from the central engine and is thus large-scale
ordered, then weak polarization evolution is expected due to the
significant polarization background (the degree is $\sim 0.5$). One
of our goals to calculate the polarization of Phase-I is to see
whether it's possible, via X-ray polarimetry, to distinguish between
the high latitude emission model and the dying central engine model.
This goal is not easy to fulfill in view of the above comparison.
However, our calculation does support that strong polarization
evolution is likely to accompany Phase-I, the decline phase of X-ray
flares (Phase-V) and possibly the late part of Phase-II. Fruitful
novel features are expected to appear in GRB X-ray afterglow
polarimetry, though detection of them is beyond the capability of
current monitors. We have witnessed the breakthrough made by {\it
Swift} XRT and we are likely to experience it once more in the
coming X-ray polarimetry era. One byproduct of the future X-ray
polarimetry is to probe the cosmological birefringence effect that
arises in some quantum gravity models (Fan, Wei \& Xu 2007 and the
references therein).

Finally we'd like to point out that the optical flares in a few
bursts (e.g., B\"oer et al. 2006)  may also have a central engine
origin (Wei, Yan \& Fan 2006). Interesting polarization signatures
are also expected.

\section*{Acknowledgments}
We thank the referee for her/his constructive comments. We'd like to
thank J. Gorosabel, J. Hjorth, J. Fynbo and J. Sollerman for
valuable comments, and B. Zhang and X. F. Wu for communication. This
work is supported by the National Science Foundation (grant
10673034, 10621303) of China, the National Basic Research Program of
China (973 program 2007CB815404), and a special grant of Chinese
Academy of Sciences (Y.Z.F and D.M.W) and by a (postdoctoral) grant
from the Danish National Science Foundation (Y.Z.F). DX is at the
Dark Cosmology Centre funded by the Danish National Research
Foundation.

\begin{appendix}
\section{Derivation of Equation (3) in this work}
\begin{figure}
\begin{picture}(0,200)
\put(100,0){\includegraphics{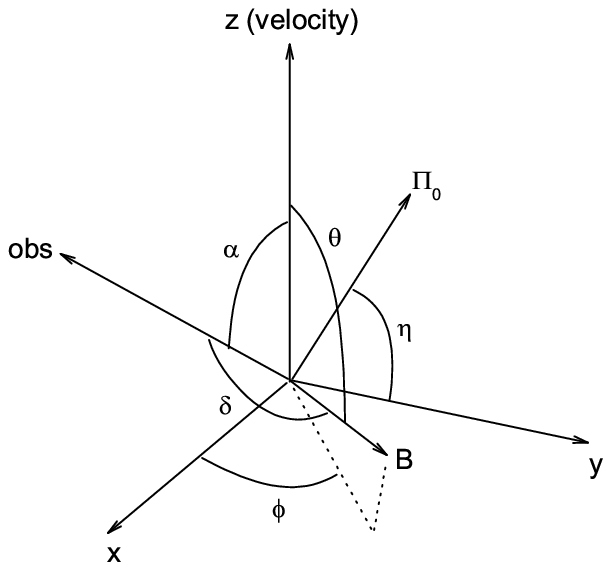}}
\end{picture}
\caption{The coordinate system to optimize the derivation (see also
Sari 1999). } \label{point}
\end{figure}

Obviously, first we need to establish a suitable coordinate system
to optimize the derivation. As stated in Sari (1999), at any point
in the shock front there is a preferred direction, the
radial direction, in which the fluid moves. We call this the parallel direction and choose the $z$%
-direction of the fluid local frame coordinate system to be in that
direction. The two perpendicular directions $(x,y)$ are assumed
equivalent, i.e., the system is isotropic in the
plane perpendicular to the direction of motion. We chose the $x$-direction to be in the plane that contains the $z$%
-direction and the direction towards the observer $\widehat{n}$, as
shown in Figure \ref{point}. Suppose now that the magnetic field has
spherical coordinates $(\theta ,\varphi )$ in that frame. A quite
general description of the distribution of the magnetic field in
such anisotropic system would be to allow different values of the
magnetic field as function of the inclination from the preferred
direction $B=B(\theta )$ as well as a probability function for the
magnetic field to be in each given inclination $f(\theta )$.

The relevant component of the magnetic field is that perpendicular
to the observer i.e. $B\sin (\delta )$, where $\delta $ is the angle
between the direction of the magnetic field and the observer. This
will produce polarization $\Pi _{0}$ in the direction perpendicular
both to the observer and to the magnetic field i.e. in the direction
$\hat{n}\times \hat{B}$. However, this polarization should be
averaged due to contributions from magnetic fields oriented
differently. On the other hand, because light is a kind of
transverse wave, its polarization should be perpendicular to its
propagation. Therefore, we project the polarization to $\hat{y}$
(i.e., the component $Q$) and $\hat{n}\times \hat{y}$ (i.e., the
component $U$), respectively.

Assuming that the emission is proportional to a certain power of the
magnetic field, $\propto B^{\epsilon }$, the total polarization from
a point-like region is then\footnote{where we have taken into
account the fact that $\hat{y}$, $\hat{n}\times \hat{y}$ and
$\hat{\Pi}=\hat{n}\times \hat{B}$ are all on the same plane which is
perpendicular to $\hat{n}$.}
\begin{eqnarray}
Q=\Pi _{0}{\int \cos (2\eta )[B(\theta )\sin \delta ]^{\epsilon
}f(\theta )\sin \theta d\phi d\theta },
\end{eqnarray}
and
\begin{eqnarray}
U=\Pi _{0}{\int \sin (2\eta )[B(\theta )\sin \delta ]^{\epsilon }f(\theta )\sin \theta d\phi
d\theta},
\end{eqnarray}
and
\begin{eqnarray}
I=\Pi _{0}{\int [B(\theta )\sin \delta ]^{\epsilon }f(\theta )\sin \theta d\phi d\theta}.
\end{eqnarray}

Since $\hat{n}=(\sin \alpha,~0,~\cos\alpha)$ and $\hat{B}=(\sin \theta \cos\phi,~\sin \theta
\sin \phi,~\cos\theta)$, we have
\[
\hat{\Pi}=\hat{n}\times \hat{B}=(-\cos \alpha \sin \theta \sin
\phi,~\cos \alpha \sin \theta \cos\phi-\sin \alpha \cos \theta,~\sin
\alpha \sin \theta \sin \phi).
\]
With $\hat{y}=(0,~1,~0)$, we obtain
\[
\cos \eta ={\hat{\Pi} \cdot \hat{y} \over |\hat{\Pi}|} =\frac{\cos
\alpha \sin \theta \cos\phi-\sin \alpha \cos \theta}{[\sin^2 \theta
\sin^2 \phi+(\cos \alpha \sin \theta \cos\phi-\sin \alpha \cos
\theta)^2]^{1/2}},
\]
and
\[
\sin \eta =\frac{\sin \theta \sin \phi}{[\sin^2 \theta \sin^2
\phi+(\cos \alpha \sin \theta \cos\phi-\sin \alpha \cos
\theta)^2]^{1/2}}.
\]
From the forms of $\cos \eta$ and $\sin \eta$ , we have
\begin{eqnarray}
\cos (2\eta)=2\cos^2 \eta-1 = \frac{(\cos \alpha \sin \theta
\cos\phi-\sin \alpha \cos \theta)^2-\sin^2 \theta \sin^2
\phi}{\sin^2 \theta \sin^2 \phi+(\cos \alpha \sin \theta
\cos\phi-\sin \alpha \cos \theta)^2},
\end{eqnarray}
and
\begin{eqnarray}
\sin (2\eta)=2\sin \eta \cos \eta = \frac{2\sin \theta \sin \phi
(\cos \alpha \sin \theta \cos\phi-\sin \alpha \cos \theta)}{\sin^2
\theta \sin^2 \phi+(\cos \alpha \sin \theta \cos\phi-\sin \alpha
\cos \theta)^2}.
\end{eqnarray}

Now substitute equation (A5) into equation (A2), we get
\begin{eqnarray}
U &=& \Pi _{0}{\int \sin (2\eta )[B(\theta)\sin \delta ]^{\epsilon
}f(\theta )\sin \theta d\phi d\theta}\nonumber\\
&\propto & \int \sin^2 \theta \sin \phi (\cos \alpha \sin \theta
\cos\phi-\sin \alpha \cos \theta) |\hat{\Pi}|^{\epsilon-2} f(\theta)
d\theta d\phi \nonumber\\
&\propto & \int^\pi_0 \sin^2\theta f(\theta)d\theta \int^{2\pi}_0
{\cal F}(\phi,\alpha,\theta) d\phi.
\end{eqnarray}
where ${\cal F}(\phi,\alpha,\theta) \equiv {\sin \phi (\cos \alpha \sin \theta \cos\phi-\sin
\alpha \cos \theta)\over [\sin^2 \theta \sin^2 \phi+(\cos \alpha \sin \theta \cos\phi-\sin
\alpha \cos \theta)^2]^{(2-\epsilon)/2}}$, and we have taken $\sin \delta=|\hat{n}\times
\hat{B}|=|\hat{\Pi}|$.

To calculate the integration for ${\cal F}(\phi,\alpha,\theta)$, we consider
\begin{equation} \int^{2\pi}_0 {\cal
F}(\phi,\alpha,\theta)d\phi=[\int^{\pi/2}_0+\int^{\pi}_{\pi/2}
+\int^{3\pi/2}_{\pi}+\int^{2\pi}_{3\pi/2}]{\cal F}(\phi,\alpha,\theta) d\phi.
\end{equation}

In regions of $\phi \in (0,~\pi/2)$ and $\phi \in (3\pi/2,~2\pi)$,
$\sin \phi$ changes the sign but $\cos \phi$ does not. So in the
above integral the contribution from the angle $\phi$ has been
cancelled out by that from $2\pi-\phi$. Similar results hold for the
contribution of regions of $\phi \in (\pi/2,~\pi)$ and $\phi \in
(\pi,~3\pi/2)$. It is then straightforward to get
\begin{equation} U\propto \int^{2\pi}_0 {\cal
F}(\phi,\alpha,\theta)d\phi=[(\int^{\pi/2}_0+\int^{2\pi}_{3\pi/2})
+(\int^{\pi}_{\pi/2}+\int^{3\pi/2}_{\pi})]{\cal
F}(\phi,\alpha,\theta) d\phi=0.
\end{equation}

{\it We thus have proved that because of the isotropy of the magnetic field in the $(x,y)$
direction, the polarization of radiation emitted from a point-like region (after averaging on
magnetic field orientation) must be in the direction perpendicular to the $z$-axis and to the
observer, i.e., in the $\hat{y}$ direction.}\\

In the case of $\epsilon=2$, both $Q$ and $I$ can be analytically
integrated. With the relations $\int^{2\pi}_0 \sin^2\phi d\phi
=\int^{2\pi}_0 \cos^2 \phi d\phi=\pi$, we have
\begin{eqnarray}
Q &\propto & \int^{\pi}_0 f(\theta)\sin\theta d\theta \int^{2\pi}_0
B^2 [(\cos \alpha \sin \theta \cos\phi-\sin \alpha \cos
\theta)^2-\sin^2
\theta \sin^2 \phi]d\phi\nonumber\\
&=& \pi \sin^2\alpha \int^\pi_0 B^2 f(\theta)\sin\theta [2\cos^2
\theta-\sin^2 \theta] d\theta,
\end{eqnarray}
Note that $<B_\parallel^2>=\int^\pi_0 B^2 f(\theta) \sin \theta
\cos^2 \theta d\theta$ and $<B_\perp^2>=\int^\pi_0 B^2 f(\theta)
\sin^3 \theta d\theta$, we have
\[
Q\propto 2 \pi \sin^2\alpha (<B_\parallel^2>-{<B_\perp^2>\over 2}).
\]
Similarly, \[I=\int^{\pi}_0 f(\theta)\sin^2 \theta B^2 |\hat{\Pi}|^2 d\theta=2\pi(\sin^2\alpha
<B^2_\parallel>+(1+\cos^2 \alpha)<B^2_\perp>/2).\]

Therefore, the net polarization is given by (see also Gruzinov 1999 and Sari 1999)
\begin{equation}
P_{\rm point}={Q \over I}=\Pi_0 {\sin^2\alpha
(<B_\parallel^2>-{<B_\perp^2>/2}) \over \sin^2\alpha
<B^2_\parallel>+(1+\cos^2 \alpha)<B^2_\perp>/2}.
\end{equation}

In the high latitude model for the X-ray sharp decline, the emission is from an angle $\alpha$
(or, equivalently $\rho$). As shown in Ghisellini \& Lazzati (1999), it is convenient to write
the polarization vector as a complex number $P=P_{\rm point}{\rm e}^{\rm 2i \theta_p}$ and
integrate it between $\psi_1$ and $2\pi-\psi_1$. Here $\theta_{\rm p}$ is the position angle
of the linear polarization in the observer frame. The polarization of a generic ring is thus
\begin{equation}
\Pi(\rho)={\int^{2\pi-\psi_1}_{\psi_1} {P_{\rm point} I d\theta_{\rm
p}}\over \int^{2\pi-\psi_1}_{\psi_1} I d\theta_{\rm p}}
=\Pi_0{(2\zeta^2-1)\sin^2 \alpha \over 2 \zeta^2 \sin^2 \alpha
+(1+\cos^2 \alpha)}{\mid \sin(2\psi_1) \mid \over 2\pi-2 \psi_1},
\end{equation}
where $\zeta^2=<B_\parallel^2>/<B_\bot^2>$.

Till here, we thus have proved equation (\ref{eq:Main}) in this work.

\end{appendix}

\end{document}